\documentstyle[12pt,epsfig]{article}
\begin{document}
\baselineskip 3.9ex
\def\be{\begin{equation}}
\def\ee{\end{equation}}
\def\bea{\begin{eqnarray}}
\def\eea{\end{eqnarray}}
\def\bd{\begin{displaymath}}
\def\ed{\end{displaymath}}
\def\l#1{\label{eq:#1}}
\def\eqn#1{(~\ref{eq:#1}~)}
\def\no{\nonumber} 
\def\av#1{{\langle  #1 \rangle}}
\title{ Distribution of Transverse Distances in Directed Animals\\
\vspace{0.2cm}
            Sumedha\thanks{sumedha@theory.tifr.res.in}~ and Deepak Dhar\thanks{ddhar@theory.tifr.res.in}\\
\vspace{0.2cm} 
Department Of Theoretical Physics\\ Tata
Institute Of Fundamental Research\\
Homi Bhabha Road, Mumbai 400005\\
India} 
\date{}
\maketitle
\begin{abstract}

	We relate $\phi(\textbf{x},s)$, the average number of sites at a
transverse distance $\textbf{x}$ in the directed animals with $s$ sites
in $d$ transverse dimensions, to the two-point correlation function
of a lattice gas with nearest neighbor exclusion in $d$ dimensions.  
For large $s$, $\phi(\textbf{x},s)$ has the scaling form $\frac{s}{R_s^d}
f(|\textbf{x}|/R_s)$, where $R_s$ is the root mean square radius of gyration of animals of $s$ sites. We determine the exact
scaling function for $d =1$ to be
$f(r) = \frac{\sqrt{\pi}}{2 \sqrt{3}}\mbox{erfc}(r/\sqrt{3})$. We also show that
$\phi(\textbf{x}=0,s)$ can be determined in terms of the animals number
generating function of the directed animals.

\end{abstract} 

The directed animals(DA) problem describes the large-scale
geometrical structure of subcritical directed percolation clusters,
and qualitatively models diverse situations such as trees, river
networks and dilute polymers in a flowing solvent \cite{green,
lubensky}. Formally, it corresponds to $p \rightarrow 0$ limit of the
directed percolation problem \cite{stauffer}. The imposed
directionality changes the universality class, and the large-scale
structure of directed animals is different from isotropic animals.
The enumeration of directed site animals in $d + 1$ dimensions is
related to hard-core lattice gas(HCLG) at negative activity with
repulsive interactions in $d$ dimensions and the Yang-Lee edge
problem in $d$ dimensions \cite{cardy,dhar1,lai,brydges}.  This
equivalence helps to obtain several exact results for the directed
animals problem \cite{dhar2,nadal,viennot,conway}. In particular, the
critical exponents $\theta$ and the transverse size exponent
$\nu_{\perp}$ are known exactly in $ d = 0, 1 $ and $2$ \cite{dhar2}.
For reviews of available exact results, see \cite{melou}.  The
problem is also related to a recently studied model of quantum
gravity \cite{difranscesco}, and there is an unexpected relation
between the number of distinct eigenvalues for Potts model partition
function on strips of width $w$ and the number of directed animals
with $w$ sites in two dimensions \cite{shrock}.

In this paper, we extend the known relation between the directed
site animals enumeration (DSAE) problem to hard-core lattice gases at negative
activity to obtain the average number of sites at a given transverse
distance $\textbf{x}$ from the origin for $d+1$-dimensional directed
animals from the density-density correlation function of the
lattice gas in $d$ dimensions. For $d=1$, using the exact generating
function for this correlation function, we determine the scaling form
for the average number of sites at a given transverse distance in a
$2$-dimensional directed animal having $s$ sites, for large $s$. In
this case, the average transverse size of the animal scales as
$s^{\nu_{\perp}}$, where $\nu_{\perp}$ is known to be $\frac{1}{2}$
for $d=1$. For large $s$, the average value of $q$-th transverse
moment $<|x|^q>$ varies as $R_s^q C_q$ , where $R_s$ is the root mean square radius of gyration of animals of $s$ sites. Using the exact scaling function we are able to determine 
the universal constants $C_q$ for all $q$.

\begin{figure}
\begin{center}
\epsfig{figure=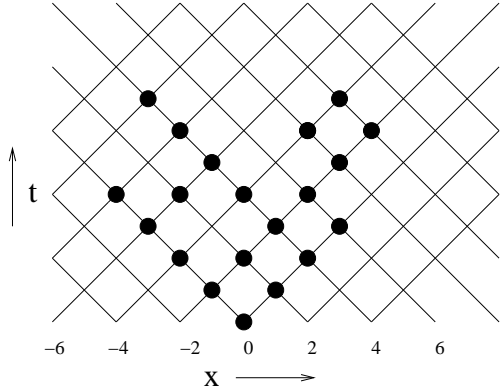,width=7cm}
\end{center}
\caption{Schematic figure of a Directed Animal of size 20 on a square lattice }
\end{figure}

We consider directed animals on a square lattice. Each site $(x,t)$
has two bonds directed outwards towards the sites $(x+1,t+1)$ and
$(x-1,t+1)$ (Fig. 1). A directed site animal $\mathcal{A}$ rooted at
the origin is a set of occupied sites including origin, such that for
each occupied site $(x,t)$ other than the origin, at least one of the
two sites $(x-1,t-1)$ and $(x+1,t-1)$ is also occupied. Fig. 1 shows a
directed animal on a square lattice in two dimensions. The
number of sites in ${\mathcal{A}}$ will be denoted by
$|\mathcal{A}|$ or $s$.  We define $n(x|{\mathcal{A}})$ as the number of
sites of $\mathcal{A}$ having the transverse coordinate $x$. In Fig. 1,
$n(x|\mathcal{A})$ takes the values $1,3$ and $0$ for $x$ equal to
$-4, 2, 5$ respectively, with $|{\mathcal{A}}|= 20$. We shall define the radius of gyration of ${\mathcal{A}}$ as $\sqrt{\frac{1}{s} \sum x^2}$, where the sum is over all sites of the animal. The animal shown in
Fig. 1, has a squared radius of gyration $21/4$.

We define $A(y)$ as  $y^{|{\mathcal{A}}|}$ summed over all animals
${\mathcal{A}}$ rooted at the origin. If $A_s$ is the number of 
distinct animals having $s$ sites, it is easy to see that 
\begin{equation}
A(y) = \sum_{\mathcal{A}} y^{|{\mathcal{A}}|} = \sum_{s=1}^{\infty} 
A_s y^s
\end{equation}

For large $s$, $A_s$ varies as $\lambda^s s^{-\theta}$, where
$\theta$ is a critical exponent. The radius of gyration 
$R_s$ is expected to vary as $ s^{\nu_{\perp}}$,
where the exponent $\nu_{\perp}$ is related to the animals number
exponent $\theta$ by the hyperscaling relation $\theta = d
\nu_{\perp}$ \cite{family}.

For a directed animal of size $s$ we define the generating function $\Psi(\textbf{x},y)$ as

\begin{equation} 
\Psi(\textbf{x};y) = \sum_{\mathcal{A}} n(\textbf{x}|{\mathcal{A}})  
y^{|{\mathcal{A}}|}
\end{equation}
where the summation over $\mathcal{A}$ is the summation over all
animal configurations. This can be written as

\begin{equation}
\Psi(\textbf{x};y) = \sum_{s} \phi(\textbf{x},{s}) A_s~ y^s
\end{equation}
where $\phi(\textbf{x},{s})$ is the value of $n(\textbf{x},{\mathcal{A}})$
averaged over all animals ${\mathcal{A}}$ of size $s$.

The DSAE problem in $d+1$-dimensions is related to time development
of thermal relaxation of a hard core lattice gas (HCLG) with nearest
neighbor exclusion on $d$ dimensional lattice \cite{dhar1}. On a $d$ dimensional body-centered hypercubical lattice (a $d$ dimensional body-centered hypercubical lattice is a hypercubical lattice having $2^{d-1}$ bonds going outward from each site), the dynamics of
the lattice gas is defined as follows: The evolution is stochastic
discrete -time Markovian.  At odd(even)  times, all the odd(even)  
sites are examined in parallel, and if a site has all neighbors
empty, its occupation number is set to $1$ with probability $p$, and
to $0$ with probability $(1-p)$. The rates of
transitions satisfy detailed balance condition corresponding to the Hamiltonian

\begin{equation}
H = +\infty \sum_{<ij>} {n_i n_j} - (\mbox{ln}z) \sum_{i} { n_i}   \label{eq:ham}
\end{equation}
with $ z = p/(1-p)$.

Let $\rho(p)$ be the average density of particles in the steady state 
of this system. In \cite{dhar1}, it was shown that we have 
\begin{equation}
A(y) = - \rho(p = -y)
\label{EqA}
\end{equation}

For the $d=1$ nearest-neighbor-exclusion lattice gas, it is straight forward to determine 
the average density  corresponding to chemical activity $p/(1-p)$.
This for square lattice gives
\begin{equation}
A(y) = -\frac{1}{2} \left[1-\sqrt{\left(\frac{1+y}{1-3 y}\right)}\right] \label{ano}
\end{equation}

The animal numbers  $A_s$, which are the coefficients in 
the Taylor expansion of $A(y)$ for square lattice can be written as
\cite{dhar2} 

\begin{equation}
A_{s} = \int_0^{2 \pi}\frac{ d\theta}{2 \pi} (1+cos\theta) 
(1+2cos\theta)^{{s}-1}. \label{ano1}
\end{equation}

For large $s$, $A_s  \sim \frac{1}{\sqrt{3 \pi}} 3^{s} 
{s}^{-\frac{1}{2}}$, which shows that in this case $\lambda =3$, and 
$\theta = 1/2$.

The derivation of Eq.(\ref{EqA}) is easily generalized to the case 
where 
the values of $p$ at different sites are different. Let the 
probability that site $i$ gets occupied at time $t+1$ given that all 
its neighbors are unoccupied at time $t$ be $p_i$. Then the rates of 
this process still satisfy detailed balance condition corresponding 
to the hamiltonian

\begin{equation}
H = +\infty \sum_{<ij>} {n_i n_j} -  \sum_{i} (\mbox{ln}z_i) { n_i}   
 \label{hamin}
\end{equation}
where $z_i = p_i/(1 - p_i)$. The probability that site
$i$ is occupied in the steady state depends on the $p_j$'s for all
sites $j$, and will be denoted by $\rho_i(\{p_j\})$. In the
corresponding DA problem, we have to define the weight of an animal
${\mathcal{A}}$ as product of weights of all occupied sites, the
weight corresponding to a site with $\textbf{x}$-coordinate $j$ being $y_j$.
Then define $A_i(\{y_j\})$ be the sum of weights of all animals
rooted at $i$. Then clearly $A_i(\{y_j\})$ is a formal power series
in the variables $\{y_j\}$.  If all $y_j = y$, this becomes
independent of $i$, and reduces to the function $A(y)$. For unequal 
$y_j$'s, Eq.(\ref{EqA}) becomes

\begin{equation}
A_i(\{y_j\}) = - \rho_i(\{p_j = - y_j\})
\label{Ain}
\end{equation}
  
Applying the operator $y_{\bf{x}} \frac{\partial}{\partial y_{\bf{x}}}$ on the 
weight of any particular animal gives us the weight multiplied by the 
number of occupied sites with transverse coordinate $\textbf{x}$ in the animal. Thus, 
clearly, we get
\begin{equation}
y_{\bf{x}} \frac{\partial}{\partial y_{\bf{x}}} A_0(\{y_j\})|_{\{ y_j=y\}} 
= 
\Psi(\textbf{x};y)
\end{equation}

Let $\Omega(\{z_j\})$ be the grand partition function for the HCLG 
given by the Hamiltonian in Eq. (\ref{hamin}) with $z_j = p_j/(1 - p_j)$. Then, $\Omega(\{z_j\})$ is a  
linear function of each of the variables $z_j$. Let $\eta_j$ be the 
indicator variable taking value $1$ if the site $j$ is occupied, and 
$0$ if not. The density-density correlation function of the gas $G(\bf{i},\bf{k})$ is 
defined as
\begin{equation}
G({\bf{i},\bf{k}};\{z_j\}) = < \eta_{\bf{k}} \eta_{\bf{i}}> -<\eta_{\bf{k}}> <\eta_{\bf{i}}> = 
z_{\bf{k}}\frac{\partial 
}{\partial z_{\bf{k}}} \rho_{\bf{i}}(\{z_j\}) .
\end{equation}

When $z_j = z$ for all $j$, the correlation function depends only on $(\bf{i}-\bf{k})$ and hence can be 
written as $G({\bf{i}-\bf{k}};z)$.

Now, using Eq.(\ref{Ain}), we get

\begin{equation}	
\Psi(\textbf{x};y) = - \frac{1}{1+y} G\left(\textbf{x};z = -\frac{y}{1+y}\right)	\label{main}
\end{equation}	

This equation is the main result of this 
paper, and equates the pair correlation function $G(\textbf{x};z)$ of the HCLG 
with the function $\Psi(\textbf{x};y)$ which gives the density profile of 
the DA problem.

For the special case $d=1$, it is a simple excercise to calculate 
$G(x;z)$ explicitly, using transfer matrix methods 
\cite{kurtz}. This gives
\begin{equation}
G(x;z) = \frac{z}{1+4 z} {\left[\frac{1-\sqrt{1-4 z}}{1+\sqrt{1+4 z}}\right]}^{|x|} \label{Gexact}
\end{equation}
and we get the explicit expression for $\Psi(x;y)$ on square lattice to be

\begin{equation}
\Psi(x;y) = \frac{y}{(1+y) (1-3y)} \left[{1 -
\sqrt{\frac{1-3 y}{1+y}}}\right]^{|x|} \left[{1 +
\sqrt{\frac{1-3 y}{1+y}}}\right]^{-|x|}
\end{equation}

This determines the density profile in the constant fugacity 
ensemble, where an animal having $s$ sites has weight $y^s$. However, 
it is more instructive to look at the profile in the constant-$s$ 
ensemble. This is obtained by looking at the Taylor coefficient of 
$y^s$ in the above equation.  This can become rather messy. However, 
the behavior for large $s$ can be determined easily.

In general, for any dimension $d$ , for large $s$, $\phi(\textbf{x},s)$ has the scaling form

\begin{equation}
\phi(\textbf{x},s) \sim \frac{s}{{R_s}^d} f\left(\frac{|\textbf{x}|}{R_s}\right)
\end{equation}

Usually the argument of the scaling function is defined only upto a multiplicative constant. We have made specific choice for this by using the variable as $|\textbf{x}|/R_s$. The normalization of scaling function $f(r)$   is chosen such that is satisfies

\begin{eqnarray}
\label{scale} \int_{-\infty}^{\infty} d^d \textbf{x} f(|\textbf{x}|) = 1\\
\int_{-\infty}^{\infty} d^d \textbf{x}~ |\textbf{x}|^2 f(|\textbf{x}|) = 1
\end{eqnarray}

 In the DA problem, as the number of animals grow as $\lambda^s$, 
the series expansions for  $\Psi(\textbf{x};y)$ or $ A(y)$ in powers of 
$y$  converge for $y <  y_c = 1/\lambda$. For $y$ near $y_c$, the 
singular part of the function $A(y)$ varies as $( 1 - y 
\lambda)^{\theta -1}$. For the HCLG problem, this corresponds to a
singularity in the series for the density $\rho$ in powers 
of activity for the activity $ z_{LY} = -( 1 + 
\lambda)^{-1}$ [Eq. (\ref{main})]. This singularity  on the negative real line is the 
Lee-Yang edge singularity for this problem.

For $z$ near  the critical value $z = z_{LY}$, this correlation 
function $G(\textbf{x};z)$  is
expected to have the scaling form

\begin{equation}
G(\textbf{x}; z = z_{LY} e^{-\epsilon}) = c~\epsilon^{-a}
g( b|\textbf{x}|\epsilon^{\nu}) + {\rm~ higher ~order ~terms~ in~ \epsilon}
\label{taylor}
\end{equation}
 where $b$ and $c$ are non-universal, lattice dependent constants. We choose $b$ such that $g(\xi) = exp(-\xi)$ for large $\xi$, and  $c$ is fixed by requiring $g(0)=1$. 

The scaling function $g(\xi)$ tends to  a constant limiting value as 
$\xi$
tends to zero, and decreases to zero exponentailly fast as $\xi$ tends to
infinity. Let the power-series expansion of  $g(\xi)$ about  $\xi = 
0$ be given
by

\begin{equation}
g(\xi) = \sum_{k=0}^{\infty} g_k \xi^k   \label{power}
\end{equation}

Substituting this in Eq.(\ref{taylor}), we get

\begin{equation}
G(\textbf{x};z) \sim \sum_{k=0}^{\infty} g_k {b}^k |\textbf{x}|^k {\left( 1-
\frac{z}{z_{LY}}\right)}^{ \nu k - a}
\end{equation}

To get $\phi(\textbf{x},s)$, we need to determine the coefficient of $z^s$
in the above expansion. From the binomial expansion of $(1 -
\frac{z}{z_{LY}})^{\nu k - a}$ we immediately get

\begin{equation}
G(\textbf{x};z) \sim \sum_{s=0}^{\infty}
 {\left(\frac{z}{z_{LY}}\right)}^s \sum_{k=0}^{\infty} g_k {b}^k |\textbf{x}|^k 
\frac{\Gamma( a+ s -\nu k)}{\Gamma(s+1) \Gamma(a - \nu k)} \label{series}
\end{equation}

For fixed $k$, for large $s$,  we have
\begin{equation}
\frac{\Gamma(s + a - \nu k)}{\Gamma( s+1)} \rightarrow  s^{a- 1 - \nu k}
\end{equation} 

Using this in Eq.(\ref{series}), we get
\begin{equation}
f(|\textbf{x}|s^{-{\nu}_{\perp}}) = \sum_{k=0}^{\infty} g_k {\left(\frac{|\textbf{x}|}{s^\nu}\right)}^k
\frac{1}{\Gamma(a - \nu k)} \label{function}
\end{equation}
where the correlation length exponent $\nu$ for the HCLG problem is the same as the transverse size 
exponent ${\nu}_{\perp}$ for directed animals.

For $d=1$, it is easy to determine $G(x;z)$ explicitly. In this case, using Eq.(\ref{Gexact}) we get the scaling form for $G(x;z)$ as
\begin{equation}
G(x;\epsilon) = \frac{1}{4 \epsilon} \mbox{exp}(-2 x \sqrt{\epsilon})
\end{equation}
so that $a = 1$, $\nu = 1/2$, $b = 2$ and $c=1/4$. The scaling function $g(\xi)$ is simply
given by 

\begin{equation}
g(\xi) = \mbox{exp}[-\xi]
\end{equation}

hence $g_k = 1/{\Gamma[k+1]}$. Using this, and the values of $a$ and $\nu$ in Eq.(\ref{function}) we get the leading singular behavior of  $\phi(x,{s}) A_s$ on square lattice to be
\begin{equation}
\phi(x,{s}) A_s = \frac{1}{4} y_c^{-s} \sum_{k=0}^{\infty} \frac{(-
\frac{\sqrt3 x}{\sqrt{s}})^k}{\Gamma(k+1)} \frac{1}{\Gamma(1-\frac{k}{2})}
\end{equation}

Since $\Gamma(1-\frac{k}{2})$ has poles when $k$ is an even integer,  only
the odd terms contribute to the sum. The resulting series is easy to
sum explicitly, giving 

\begin{equation}
\phi(x,{s}) A_s = \frac{1}{4} 3^{s} \mbox{erfc}\left(\frac{\sqrt3 x}{
2 \sqrt{s}}\right)+ correction~to~scaling~terms
\end{equation}
where $\mbox{erfc}(x) = \frac{2}{\sqrt{\pi}} \int_x^{\infty} e^{-x^2} dx$.

This gives $\phi(x,s)$ for large $s$ to be

\begin{equation}
\phi(x,{s}) = \frac{\sqrt{3 \pi s}}{4} ~~\mbox{erfc}\left(\frac{\sqrt3 x}{2 
\sqrt{s}}\right) + correction~to~scaling~terms
\end{equation}

From $\phi(x,s)$ we can as well derive the expression for $q$th transverse
moment of the directed animals. The $q$th transverse moment of a cluster
of size ${s}$ denoted by $\mu_{q{s}}$ is defined as

\begin{equation}
\mu_{q{s}} = \sum_{i=-{s}}^{{s}} \phi(i,{s}) |i|^q
\end{equation}
Using the scaling form of $\phi(i,{s})$, for large ${s}$, we see that $(\mu_{qs}/s R_s^q)$ is a universal constant. Denoting it by $C_q$, we get

\begin{equation}
\frac{\mu_{qs}}{s R_s^q}= C_q = \int_{-\infty}^{\infty} d^d \textbf{x} |\textbf{x}|^q f(|\textbf{x}|)   \label{eq:trans}
\end{equation}
For $d=1$, we get 

\begin{equation}
C_q = \frac{1}{q+1} (3)^{\frac{q}{2}}\Gamma \left[1+\frac{q}{2} \right]
\end{equation}

and $R_s = \frac{2}{3} \sqrt{s}$. Hence we get the scaling function(\ref{scale}) in $d=1$ to be 
\begin{equation}
f(r) = \frac{\sqrt{\pi}}{2\sqrt3}\mbox{erfc}\left(\frac{r}{\sqrt{3}}\right)
\end{equation}

In the entire low-density phase of the HCLG, in any dimension $d$, we 
expect the 
correlation function $G(\textbf{x};z)$ to have an exponential decay at large 
$|\textbf{x}|$. But the behavior of the scaling function $f(r)$ for large 
$r$ is in general different. Suppose $\ln f(r)$ varies as 
$-r^{\alpha}$ for large $r$. Putting this behavior in Eq.(3),   
and using $A_s \sim \lambda^s$ and Eq. (\ref{main}) , we get
\begin{equation}
G(\textbf{x};z_{LY} e^{-\epsilon}) \sim \sum_s  \mbox{exp}( - {|\textbf{x}|^{\alpha} s^{-\nu 
\alpha}- \epsilon s}) 
\end{equation}
   
For large $|\textbf{x}|$, the integral can be estimated by steepest descent, and 
gives $\log g(\xi)$ varying as $\xi^{\frac{\alpha}{1 + \nu \alpha}}$. 
Since this should be linear in $\xi$, we see that $\alpha = 
\frac{1}{1 - \nu}$. As a check, we see that in $d=1$, $\nu = 1/2$, 
and $f(r)$ varies as $\mbox{exp}(- r^2)$ for large $r$.

The case $x=0$ is special, in that the density-density correlation function 
$G(0,z)$ is always equal to $\rho ( 1 - \rho)$ for hard-core 
lattice gas for any $d$ dimensional lattice. Hence, if one knows $\rho$ as a function of the activity 
$z$ ( equivalently, in the DA problem, one knows the animal numbers 
generating function $A(y)$), then one can determine $\Psi(0;y)$ in 
terms of $A(y)$ alone. For a $d+1$ dimensional DA on a body-centered hypercubical lattice it is given by 
 	
\begin{equation}
\Psi(0;y) =  \frac{1}{1+y} A(y) [1+A(y)]  \label{eq:zerox}
\end{equation} 
For $d=1$, using the $A(y)$ from Eq. (\ref{ano}), we get
for the square lattice DA problem
\begin{equation}
\Psi(0;y) = \frac{y}{1-2 y - 3 y^2}
\end{equation}
Expanding in powers of $y$ we get

\begin{equation}
\phi(0,{s}) = \frac{3^{s} + (-1)^{{s}-1}}{4 A_s}
\end{equation}
where $A_s$ is as given in Eq.(\ref{ano1}). For large ${s}$, $\phi(0,s)$ varies as $s^{1/2}$ as expected.

Similar analysis can be extended to higher dimensions. In general 
the scaling function $f(r)$ tends to a finite value as $r$ tends 
to $0$, and hence $\phi(0,s)$ varies as $s^{1 -\theta}$.

For  $d > 7$, mean-field theory becomes asymptotically exact \cite{derbez} with $R_s \sim s^{1/4}$ and the scaling form of $f(r)$ is
 
\begin{equation}
f(r) = \frac{2}{{\Omega}_d}\frac{1}{r^{d-2}} \mbox{exp}(-r^2)
\end{equation}

where ${\Omega}_d$ is the surface area of a $d$ dimensional unit sphere and is equal to ${\Omega}_d = \frac{2 {\pi}^{d/2}}{\Gamma(d/2)}$.

\end{document}